\documentclass[runningheads]{llncs}
\usepackage[T1]{fontenc}
\usepackage{graphicx}
\usepackage{amsmath,amssymb,amsfonts}

\usepackage{tikz}
\usepackage{hyperref}

\newcommand{\tikzcircle}[2][red,fill=red]{\tikz[baseline=-0.5ex]\draw[#1,radius=#2] (0,0) circle ;}%

\begin{document}
\title{A Term-Rewriting Semantics for \\ Pure Quantum States}
%
%
\author{Dan-Adrian German\orcidID{0000-0002-3746-1631} }
\authorrunning{Dan-Adrian German, Indiana University Bloomington}
%
\institute{Luddy School of Informatics, Computing and Engineering \\Indiana University, Bloomington IN 47408, USA \\
\email{dgerman@iu.edu}\\
}
\maketitle              
\begin{abstract}
In 2017, Terry Rudolph introduced an elementary rewriting system that relies on a representation of quantum states as misty states to accurately describe the basics of quantum circuits and quantum computation to high-school and middle-school students. The accessibility and effectiveness of the system are remarkable: every calculation can be done to good-enough accuracy, and perhaps with a small overhead, using just a tiny, universal set of gates chosen to take advantage of a remarkable mathematical result by Yaoyun Shi, leveraging another powerful result by A. Y. Kitaev. The misty formalism greatly simplifies calculations and makes them accessible to first-time learners using only simple arithmetic, and without sacrificing accuracy; it, too, is universal, inasmuch as you can use it to do any quantum calculation with maybe just a small overhead. We don't advocate that we should recast all of quantum theory into this formalism. The misty state picture is a good way of getting people to the heart of some nontrivial quantum theory without having to first absorb a huge amount of (what might initially seem largely) irrelevant math. Our argument is that the misty formalism can effectively be used to facilitate a transition to the full, conventional quantum-mathematical apparatus. To this end, we start by reviewing the original proposal, consider its strengths and limitations, and show it in action via entanglement swapping. We then extend the formalism through a new category of (irreducible) misty states acting as fixed points, and present the GHZ game in this new, general setting and representational semantics.

\keywords{Quantum computing \and  Rewriting systems \and Mathematics of quantum mechanics \and  Misty states formalism \and  First-time learners \and Entanglement swapping \and  GHZ game.}
\end{abstract}
\section{Introduction}

In 2017 Terry Rudolph published a slim volume entitled ``Q is for Quantum'' \cite{b2}. Soon thereafter Ken Zetie reviewed the book in The Mathematical Gazette 
\cite{b1}  and wrote: ``Terry Rudolph is a professor of physics at Imperial College, London who works and publishes in the field of quantum theory. I point
this out\footnote{Terry 
Rudolph is also one of the founders of the company PsiQuantum \cite{b3,b4,b16} where he is also chief architect having invented 
the fusion-based quantum computing approach upon which PsiQuantum's overall approach and implementation is based.}
because it would be easy to dismiss this jokey-looking and -sounding book as yet another tome with quantum in the title. It is far, far more than that.'' He
then goes on to explain what the book consists of: ``Just about every attempt at explaining quantum mechanics falls into one of two sets. Set $A$ uses the full 
mathematical power of wave mechanics, perhaps the Dirac notation, and rigorously derives results that are convincing but lacking in interpretation. Set $B$ discusses 
interpretations and ideas in quantum mechanics but shies away from anything mathematical [...]. Rudolph comes along to change this, and he does so by reinventing the 
game. [He] introduces the maths of quantum theory by talking about states which are represented by diagrams. There are simple rules about 
what happens to diagrams as they pass through systems, and these systems turn out to be elements of quantum computers. What is rigorous is that the diagrams are constructed 
based on specific rules about combinations, commas and their manipulation; these rules, although trivial to apply diagrammatically, map exactly to the rules one would use in 
creating entangled wavefunction states, etc. In other words, he invents a new mathematical tool for doing calculations in quantum theory, and it is one that a smart 12 year-old
could probably apply.'' 

The best and quickest introduction to the book is a 17 minute long video \cite{b5} by Terry Rudolph available  online\footnote{The entire book is also 
available for free at the same address.}.  
This paper is a short account on how the original system can be extended beyond the scope of the book. Extensions described here lead 
to interesting or surprising (but very accessible) results, further demonstrating the effectiveness of the approach. Eventually a genuine bridge is created to the actual mathematics 
behind quantum computation. This paper can be used by students and teachers alike but it is mainly directed to HS and middle school teachers as it addresses the underpinnings of the
proposed system. We aim to keep it  self-contained the occasional reference to \cite{b2} notwithstanding.

\section{Qubits}

The classical bits 0 and 1 will be represented as $|0\rangle$ and $|1\rangle$.
\begin{figure}[h!]
    \centering
    \includegraphics[width=0.30\textwidth]{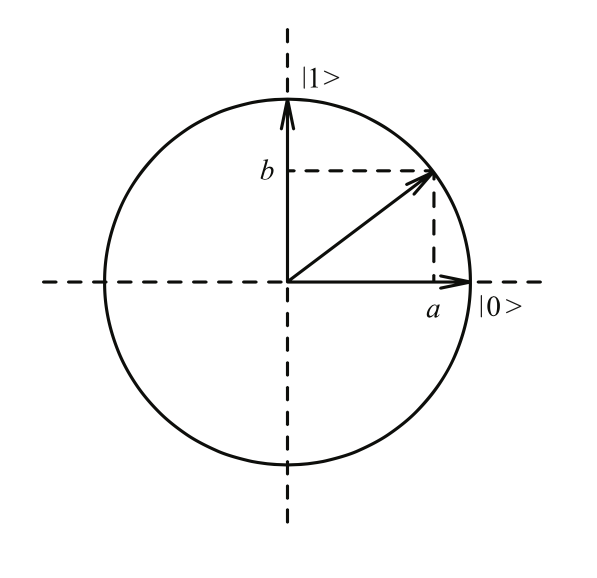}
    \caption{Qubit $|\Psi\rangle = a |0\rangle + b |1\rangle$ with real probability amplitudes $a, b \in \mathbb{R}$ expressed in the computational basis and represented on the unit circle.}
    \label{fig:001}
\end{figure}
A qubit is a linear combination of the classical bits: 
$|\Psi\rangle = a |0\rangle + b |1\rangle$. We call $a$ and $b$ 
probability amplitudes and in this paper we restrict ourselves to 
real values: $a, b \in \mathbb{R}$. 
When we measure $|\Psi\rangle$ 
we obtain either 0 or 1 with a probability that is the square of the 
corresponding probability amplitude, so: $a^2 + b^2 = 1$. 

Following 
\cite{b6}, if we draw two orthogonal axes in the plane we can take 
the unit vector on the horizontal axis to be 
$|0\rangle = \begin{pmatrix} 1 \\ 0 \end{pmatrix}$ 
and the unit vector on the vertical axis to be 
$|1\rangle = \begin{pmatrix} 0 \\ 1 \end{pmatrix}$ 
and can represent a qubit as an arrow on  the unit circle 
as shown in  Figure \ref{fig:001}: 
$\begin{pmatrix} a \\ b \end{pmatrix} = a |0\rangle + b |1\rangle$.

\section{Superposition}

We can work out a related qubit representation:

\begin{figure}[h!]
    \centering
    \includegraphics[width=0.30\textwidth]{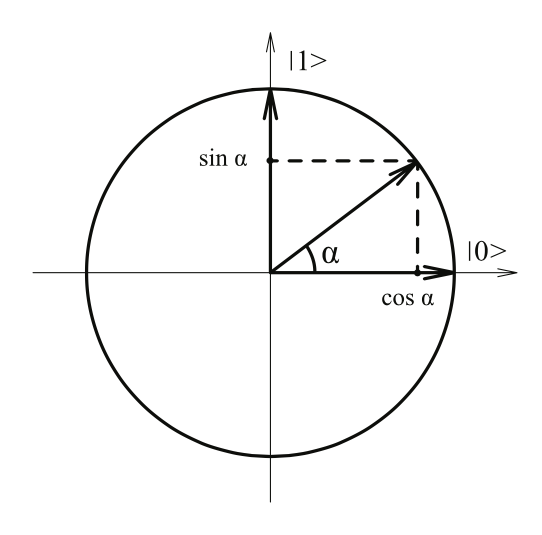}
    \caption{Same qubit $|\Psi\rangle = a |0\rangle + b |1\rangle$ represented in the complex 
    plane (isomorphic with the real plane as real vector spaces) is: $e^{i\alpha} = \cos{\alpha} + i \sin{\alpha}$}
    \label{fig:002}
\end{figure}
We now want to establish what quantum state 
results from the superposition of two quantum states 
and we first write:
$$e^{i\alpha} + e^{i\beta} = 2 \cos{\frac{\alpha - \beta}{2}} e ^ {i \frac{\alpha + \beta}{2}}$$

Through normalization we can further write: 
$$\{e^{i \alpha}, e^{i \beta}\} \rightarrow e ^{i \frac{\alpha + \beta}{2}}$$

\section{Misty States}

We represent classical bits 0 and 1 as \tikzcircle[black, fill=white]{5pt} 
and \tikzcircle[black, fill=black]{5pt}. 
As qubits 
$\tikzcircle[black, fill=white]{5pt} = \begin{pmatrix} 1 \\ 0 \end{pmatrix}$ 
and 
$\tikzcircle[black, fill=black]{5pt} = \begin{pmatrix} 0 \\ 1 \end{pmatrix}$.
A misty state is just an unnormalized quantum state\footnote{ So, it's a superposition of basis vectors and/or other states.}. In the book a misty 
state is drawn in a cloud\footnote{Although square brackets are also used 
with the letters \texttt{W} and \texttt{B} sometimes.}, here we use curly 
braces as the superposition operator. For example 
$\{ \tikzcircle[black, fill=white]{5pt}, \tikzcircle[black, fill=white]{5pt}, \tikzcircle[black, fill=black]{5pt} \}$ 
can be represented by the vector 
$\begin{pmatrix} 2 \\ 1 \end{pmatrix} $.
After normalization we have: 
$$\{ \tikzcircle[black, fill=white]{5pt}, \tikzcircle[black, fill=white]{5pt}, \tikzcircle[black, fill=black]{5pt} \} = \frac{1}{\sqrt{5}} \begin{pmatrix} 2 \\ 1\end{pmatrix} = \frac{2}{\sqrt{5}}|0\rangle + \frac{1}{\sqrt{5}}|1\rangle$$

\subsection{Coefficients}

Because misty states don't need to be normalized their formalism (as 
presented in the book) is entirely pure, that is, devoid of numbers 
(coefficients). However, given an arbitrary misty state, how can one 
determine the probability of each outcome in the state upon measurement? 
The ``squaring rule'' is the only rule with numbers in the
original (pure) misty state formalism and says:

$$\{ \underbrace{ \tikzcircle[black, fill=white]{5pt} \ldots  \tikzcircle[black, fill=white]{5pt} }_n, \overbrace{ \tikzcircle[black, fill=black]{5pt} \ldots  \tikzcircle[black, fill=black]{5pt} }^m \} = \frac{n}{\sqrt{n^2 + m^2}} |0\rangle + \frac{m}{\sqrt{n^2 + m^2}} |1\rangle$$

\subsection{Operations}

An elementary quantum operation is analogous to an elementary gate in 
a classical circuit. One of the most important examples is the Hadamard 
gate, denoted by H, which operates on a single qubit as follows: 
$$\textrm{H}(\tikzcircle[black, fill=white]{5pt}) = \{\tikzcircle[black, fill=white]{5pt},\tikzcircle[black, fill=black]{5pt}\} = \frac{1}{\sqrt{2}}(|0\rangle + |1\rangle) = |+\rangle$$
$$\textrm{H}(\tikzcircle[black, fill=black]{5pt}) = \{\tikzcircle[black, fill=white]{5pt},\overline{\tikzcircle[black, fill=black]{5pt}}\} = \frac{1}{\sqrt{2}}(|0\rangle - |1\rangle) = |-\rangle$$

Normally H (i.e., the {\small PETE} box in the book) is represented by the unitary matrix 
$\textrm{H} = \frac{1}{\sqrt{2}}\begin{pmatrix} 1 & 1 \\ 1 & -1\end{pmatrix}$. 
The definitions above effectively extract the two columns of
this matrix. The horizontal line 
above $\tikzcircle[black, fill=black]{5pt}$ represents a phase of $\pi$.

\subsection{Identities}

Suppose we are being asked to prove that 
$\textrm{H}(\textrm{H}(\tikzcircle[black, fill=black]{5pt})) = \tikzcircle[black, fill=black]{5pt}$ following the examples and methods presented in the
book. 
We could then write the following:
\begin{equation*}
\begin{split}    
\textrm{H}(\textrm{H}(\tikzcircle[black, fill=black]{5pt})) & = \textrm{H}(\{ \tikzcircle[black, fill=white]{5pt}, \overline{\tikzcircle[black, fill=black]{5pt}}\}) = \{ H(\tikzcircle[black, fill=white]{5pt}), \overline{H(\tikzcircle[black, fill=black]{5pt})} \} = \\  
 & =  \{ \{ \tikzcircle[black, fill=white]{5pt}, \tikzcircle[black, fill=black]{5pt} \}, \overline{\{ \tikzcircle[black, fill=white]{5pt}, \overline{\tikzcircle[black, fill=black]{5pt}} \} } \} =  \{ \{ \tikzcircle[black, fill=white]{5pt}, \tikzcircle[black, fill=black]{5pt} \}, \{\overline{\tikzcircle[black, fill=white]{5pt}}, \overline{\overline{\tikzcircle[black, fill=black]{5pt}}} \} \} = \\ 
 & = \{ \tikzcircle[black, fill=white]{5pt}, \tikzcircle[black, fill=black]{5pt}, \overline{\tikzcircle[black, fill=white]{5pt}}, \tikzcircle[black, fill=black]{5pt} \} = \{ \tikzcircle[black, fill=black]{5pt}, \tikzcircle[black, fill=black]{5pt} \} = \tikzcircle[black, fill=black]{5pt}
\end{split}
\end{equation*}
If in this context we are 
asked to simplify $\{ \tikzcircle[black, fill=white]{5pt}, \{ \tikzcircle[black, fill=white]{5pt}, \tikzcircle[black, fill=black]{5pt} \}\}$ we find that none
of the methods and examples in the book help and the state appears irreducible. 

\subsection{Eigenvectors}

In contrast with states that appear legitimate (well-formed) but 
irreducible we can also identify states that simply don't admit a 
representation, for example: 
$$\frac{1}{\sqrt{3}} |0\rangle + \sqrt{\frac{2}{3}}|1\rangle$$
Since misty states always contain a whole number of black 
and white balls (in the book) the state above does not admit 
a representation in the (original, let's say, pure) misty state formalism. 
This follows from the ``squaring rule'' (that appears on page 83 in the book)
and because $3$ cannot be expressed as a sum of two perfect squares.
But what's important here is that $\{ \tikzcircle[black, fill=white]{5pt}, \{ \tikzcircle[black, fill=white]{5pt}, \tikzcircle[black, fill=black]{5pt} \}\}$ is a fixed point for H,  so
(by linearity):
$$\textrm{H}(\{ \tikzcircle[black, fill=white]{5pt}, \{ \tikzcircle[black, fill=white]{5pt}, \tikzcircle[black, fill=black]{5pt} \}\}) = \{ 
\{ \tikzcircle[black, fill=white]{5pt}, \tikzcircle[black, fill=black]{5pt} \}, 
\tikzcircle[black, fill=white]{5pt} \} = 
\{ \tikzcircle[black, fill=white]{5pt}, \{ \tikzcircle[black, fill=white]{5pt}, \tikzcircle[black, fill=black]{5pt} \}\} $$
And, on second thought, this also means that $\{ \tikzcircle[black, fill=white]{5pt}, \{ \tikzcircle[black, fill=white]{5pt}, \tikzcircle[black, fill=black]{5pt} \}\}$ is an eigenvector \cite{b7} of H.
And indeed so, since: 
$$\{ \tikzcircle[black, fill=white]{5pt}, \{ \tikzcircle[black, fill=white]{5pt}, \tikzcircle[black, fill=black]{5pt} \}\} = \frac{1}{\sqrt{2}} |0\rangle + \frac{1}{\sqrt{2}}\Big(\frac{1}{\sqrt{2}} |0\rangle + \frac{1}{\sqrt{2}} |1\rangle\Big) $$

\subsection{Extensions}

Now the introduction of eigenvectors allows us to discuss a 
possible extension of the original formalism that, in this context, 
appears natural. 
So, 
Terry proposes\footnote{This is not in the book, it's from an email 
dated October 4, 2024.} that we should start by getting students used 
to the simplified notation 
$\{ a\/\tikzcircle[black, fill=white]{5pt}, b\/\tikzcircle[black, fill=black]{5pt}\}$ where $a, b \in {\mathbb N}$ are integers 
representing the number of copies. 
Although not strictly necessary one could also emphasize that 
ultimately when we calculate 
probabilities if $a, b$ share any common factors they 
can be cancelled, just like when reducing fractions to lowest form, 
because cancellation doesn’t change the probability calculations for 
what we can actually observe. 
We can then ask the interesting question: 
\begin{quote}
``Is there any mist which passes through the {\footnotesize PETE} box 
unchanged?''      
\end{quote}
And, at first glance the answer seems ``obviously not!'', because a {\footnotesize PETE} box (Hadamard gate) implements the evolution:
$$\{a\tikzcircle[black, fill=white]{5pt},b\tikzcircle[black, fill=black]{5pt}\} \rightarrow \{(a+b)\tikzcircle[black, fill=white]{5pt},(a-b)\tikzcircle[black, fill=black]{5pt}\}$$
but the equations 
\begin{equation*}
\centering
\left\{\begin{aligned}
a&=a+b \\
b &= a-b 
\end{aligned}\right.
\end{equation*}
do not have nontrivial solutions. 

\newpage 
\noindent So we need to change the question to: 
\begin{quote} 
``Is there any mist which passes through 
the {\footnotesize PETE} box (Hadamard gate)
such that the probabilities of observing 
$\tikzcircle[black, fill=white]{5pt}$ or 
$\tikzcircle[black, fill=black]{5pt}$ 
are unchanged?''
\end{quote} 
For that we have to solve
\begin{equation*}
\centering
\left\{\begin{aligned}
a^2&=\frac{(a+b)^2}{(a+b)^2+(a-b)^2} \\
b^2&=\frac{(a-b)^2}{(a+b)^2+(a-b)^2} 
\end{aligned}\right.
 \end{equation*}
and for this the ratio $\frac{a}{b}$ needs to be an irrational\footnote{At this point, in the classroom, one can tell the story of poor Hippasus.} number.

\subsection{Term-Rewriting}

Now we show how to reduce the state $\{ \tikzcircle[black, fill=white]{5pt}, \{ \tikzcircle[black, fill=white]{5pt}, \tikzcircle[black, fill=black]{5pt} \}\}$:
$$\{ \tikzcircle[black, fill=white]{5pt}, \{ \tikzcircle[black, fill=white]{5pt}, \tikzcircle[black, fill=black]{5pt} \}\} \rightarrow  \{ e^{i \cdot 0}, \{ e ^ {i \cdot 0 }, e ^ {i \cdot \frac{\pi}{2}} \}\} =  \{ e^{i \cdot 0}, e ^ {i \cdot \frac{\pi}{4}} \}  = e^{i \cdot \frac{\pi}{8}} $$
And that's exactly right since $$\{ \tikzcircle[black, fill=white]{5pt}, \{ \tikzcircle[black, fill=white]{5pt}, \tikzcircle[black, fill=black]{5pt} \}\} = \cos{\frac{\pi}{8}}|0\rangle + \sin{\frac{\pi}{8}}|1\rangle$$

\subsection{The General Case}

In the general case a state is $$|\Psi_1\rangle = \{ n_1\tikzcircle[black, fill=white]{5pt}, m_1\tikzcircle[black, fill=black]{5pt} \} \approx e^ {i \cdot \alpha_1} $$ with $\tan \alpha_1 = \frac{m_1}{n_1}$. 
We then ask if it's possible to reduce the superposition $\{ \{ n_1\tikzcircle[black, fill=white]{5pt}, m_1\tikzcircle[black, fill=black]{5pt} \}, 
\{ n_2\tikzcircle[black, fill=white]{5pt}, m_2\tikzcircle[black, fill=black]{5pt} \}
\}$ by using the rules, methods and examples in the book. And we're going to state (based on all we've seen and said thus far) that the superposition is described by 
$$\{ n\tikzcircle[black, fill=white]{5pt}, m\tikzcircle[black, fill=black]{5pt} \} \approx e^ {i \cdot \alpha}$$ such that $\alpha = \frac{\alpha_1 +\alpha_2}{2}$ and $\tan{\alpha} = \frac{m}{n} = \frac{\sin{\alpha_1} + \sin{\alpha_2}} {\cos{\alpha_1} + \cos{\alpha_2}}$. 

We can see that $\frac{m}{n} = \frac{m_1 + m_2}{n_1 + n_2}$ iff $n_1^2 + m_1^2 = n_2^2+m_2^2$. 
So wherever 
you see this transformation in the book (or further in this paper) 
please verify that we're indeed in this specific context where the simplified rule applies. One can, however, work out
the general case but the
rules (formulas) are going to be a little bit more involved and
they are not relevant to the rest of our presentation. 

\section{Examples}

\noindent
\textcircled{{\footnotesize 1}}. The misty state representation of the second eigenvector of the Hadamard gate, $H$ (or, {\footnotesize PETE} box, in the book) is 
$\{ \tikzcircle[black, fill=white]{5pt}, \overline{\{\tikzcircle[black, fill=white]{5pt}, \tikzcircle[black, fill=black]{5pt}\}}\}$. To prove this statement we
can proceed in more than one way. For example, one can try
translating the misty state suggested above directly in 
Dirac notation and then work out 
the algebra to find the expression
of the second eigenvector (if that's already known). One can
also reason with vectors as we've shown before. 
Finally it is easy
(and straightforward, also good practice) to try to prove the following
using only misty states (this is just a straightforward application of linearity):
$$H(\{ \tikzcircle[black, fill=white]{5pt},  \overline{\{ \tikzcircle[black, fill=white]{5pt}, \tikzcircle[black, fill=black]{5pt}  \} } \}) = \overline{\{ 
\tikzcircle[black, fill=white]{5pt},  \overline{\{ \tikzcircle[black, fill=white]{5pt}, \tikzcircle[black, fill=black]{5pt} \} } \} }$$

\noindent \textcircled{{\footnotesize 2}}. Consider the following Qiskit circuit:
\begin{figure}[h!]
    \centering
    \includegraphics[width=0.93\textwidth]{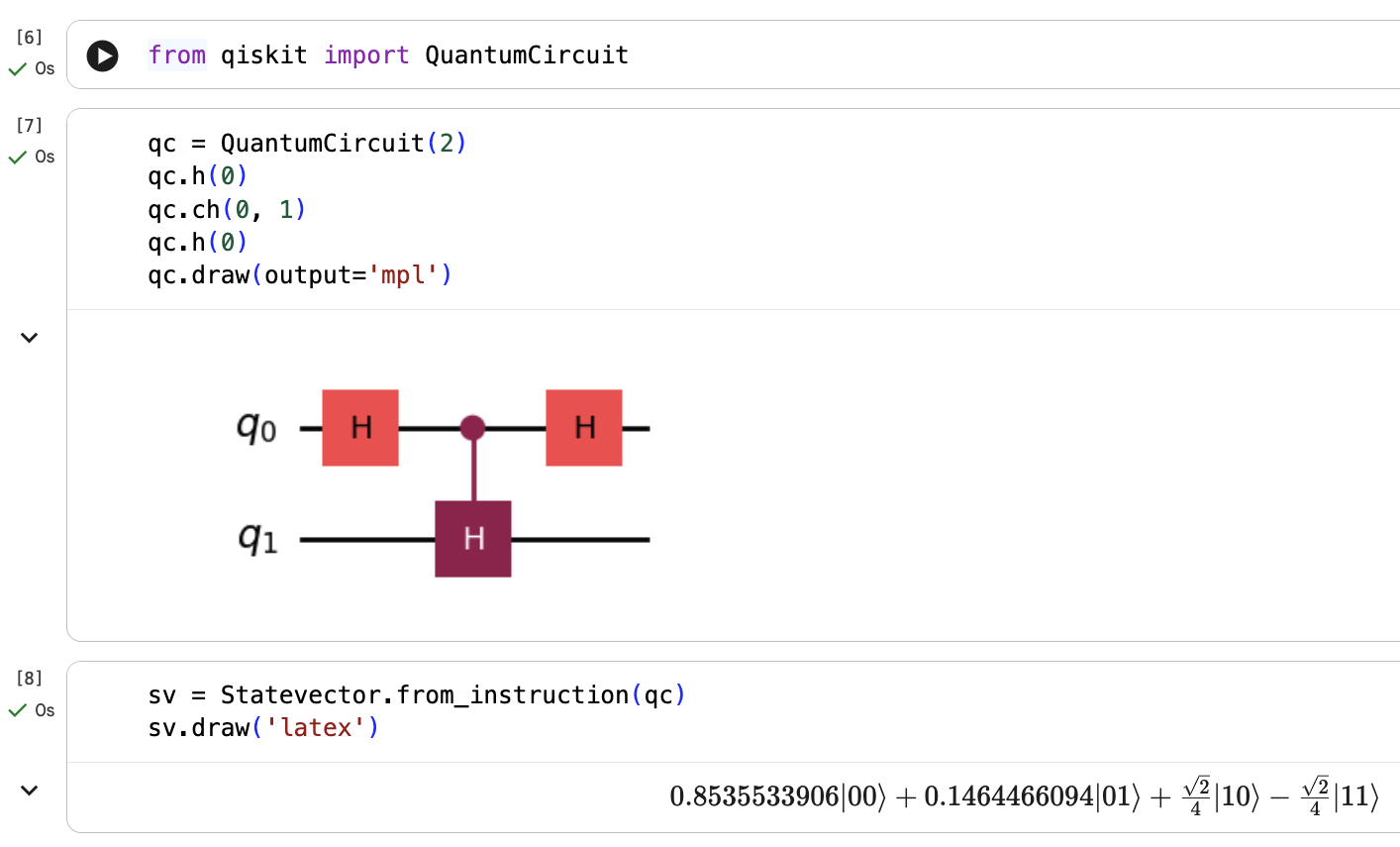}
    \caption{Qiskit circuit that produces (alternatively) both Hadamard eigenvectors.}
    \label{fig:003}
\end{figure}

To prove that a state is an eigenvector 
we simply test that it is a fixed point. The circuit presented here produces 
both eigenvectors of the Hadamard gate: we feed $|0\rangle$ in both inputs 
and then measure the first output. If the  measurement detects a 0 (as it 
does in example 3) we have one eigenvector of H on the other output line. 
If (as in example 4) we detect a 1 then the other output contains the other
eigenvector of the Hadamard gate. 
Only a certain subset of quantum gates can 
be used with the pure misty state formalism and controlled-Hadamard is not 
among those.  But we can handle easily this circuit in the extended misty 
state formalism and the derivation is simple, beautiful. 

We start 
with input $|\Psi_1\rangle =  \tikzcircle[black, fill=white]{5pt}\tikzcircle[black, fill=white]{5pt} $ and after we apply the first Hadamard gate the state 
becomes $|\Psi_2\rangle =  \{\tikzcircle[black, fill=white]{5pt}, \tikzcircle[black, fill=black]{5pt} \} \tikzcircle[black, fill=white]{5pt} = 
\{ \tikzcircle[black, fill=white]{5pt}\tikzcircle[black, fill=white]{5pt}, \tikzcircle[black, fill=black]{5pt}\tikzcircle[black, fill=white]{5pt}\}$. 
With the first qubit as control we get      
$|\Psi_3\rangle =  \{\tikzcircle[black, fill=white]{5pt}\tikzcircle[black, fill=white]{5pt}, \tikzcircle[black, fill=black]{5pt}\{\tikzcircle[black, fill=white]{5pt},\tikzcircle[black, fill=black]{5pt}\}\}$. We then 
apply a Hadamard on the first qubit and observe it. If we see $\tikzcircle[black, fill=white]{5pt}$ then we will have collapsed the second qubit to the first eigenvector of H. 
To recap, the derivation is:
\begin{equation*} \label{eq1}
\begin{split}
|\Psi_4\rangle & =  \{\textrm{H}(\tikzcircle[black, fill=white]{5pt})\tikzcircle[black, fill=white]{5pt}, \textrm{H}(\tikzcircle[black, fill=black]{5pt})\{\tikzcircle[black, fill=white]{5pt},\tikzcircle[black, fill=black]{5pt}\}\}\\
 & = \{\{\tikzcircle[black, fill=white]{5pt}, \tikzcircle[black, fill=black]{5pt} \}\tikzcircle[black, fill=white]{5pt}, \{\tikzcircle[black, fill=white]{5pt}, \overline{\tikzcircle[black, fill=black]{5pt}} \}\{\tikzcircle[black, fill=white]{5pt},\tikzcircle[black, fill=black]{5pt}\}\}\\
 & =  \{\{\tikzcircle[black, fill=white]{5pt}\tikzcircle[black, fill=white]{5pt}, \tikzcircle[black, fill=white]{5pt}\{\tikzcircle[black, fill=white]{5pt},\tikzcircle[black, fill=black]{5pt}\}\},\{\tikzcircle[black, fill=black]{5pt}\tikzcircle[black, fill=white]{5pt}, \overline{\tikzcircle[black, fill=black]{5pt}}\{\tikzcircle[black, fill=white]{5pt},\tikzcircle[black, fill=black]{5pt}\}\}\} \\
 & = \{ \tikzcircle[black, fill=white]{5pt}\{ \tikzcircle[black, fill=white]{5pt}, \{ \tikzcircle[black, fill=white]{5pt}, \tikzcircle[black, fill=black]{5pt}\}\}, \tikzcircle[black, fill=black]{5pt}\{\tikzcircle[black, fill=white]{5pt},\overline{\{ \tikzcircle[black, fill=white]{5pt}, \tikzcircle[black, fill=black]{5pt}\}}\}\}.
\end{split}
\end{equation*}

\noindent 
\textcircled{{\footnotesize 3}}. This shows how one can measure the state partially with Qiskit to 
confirm the validity of our theoretical calculations.
\begin{figure}[h!]
    \centering
    \includegraphics[width=0.93\textwidth]{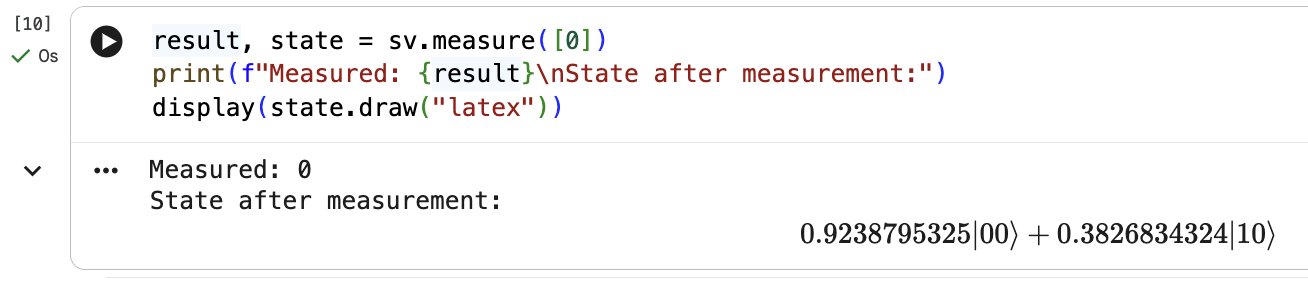}
    \caption{When the first line is a 0  the first Hadamard eigenvector is on the other line.  }
    \label{fig:004}
\end{figure}

\noindent 
\textcircled{{\footnotesize 4}}. If we measure a 1 on the first line: 
\begin{figure}[h!]
    \centering
    \includegraphics[width=0.93\textwidth]{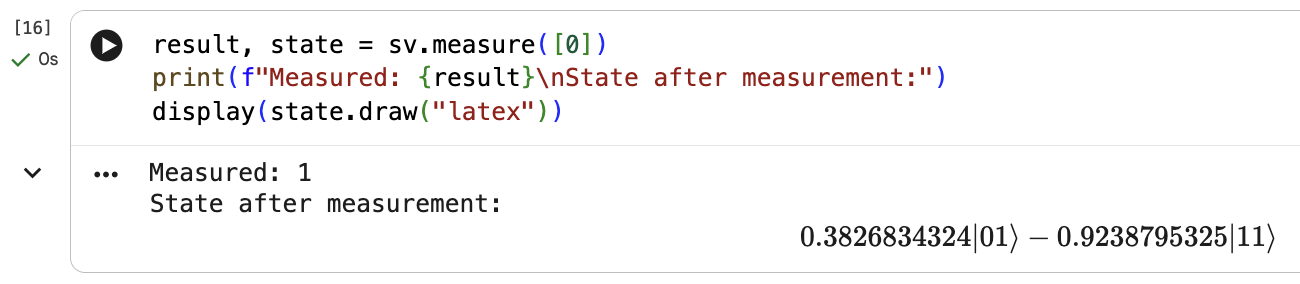}
    \caption{Getting the other Hadamard eigenvector from  our circuit.}
    \label{fig:005}
\end{figure}

\noindent Now the output of the circuit: 
$\{ \tikzcircle[black, fill=white]{5pt} \{ \tikzcircle[black, fill=white]{5pt}, \{ \tikzcircle[black, fill=white]{5pt}, \tikzcircle[black, fill=black]{5pt}\}\}, \tikzcircle[black, fill=black]{5pt} \{ \tikzcircle[black, fill=white]{5pt}, \overline{\{ \tikzcircle[black, fill=white]{5pt}, \tikzcircle[black, fill=black]{5pt}\}}\} \}$ may be easier to interpret. One can see clearly that it's a superposition of two qubit states in which the first qubit is either 0 or 1 while  the second qubit is either one, or the other of the two Hadamard eigenvectors.

\noindent \textcircled{{\footnotesize 5}}. Let's simplify the following misty state:  $$\{\{\{\tikzcircle[black, fill=white]{5pt}, \tikzcircle[black, fill=black]{5pt}\}, \tikzcircle[black, fill=black]{5pt}\} , \{\tikzcircle[black, fill=white]{5pt}, \{\tikzcircle[black, fill=white]{5pt}, \tikzcircle[black, fill=black]{5pt}\}\}\}$$
This state  reduces to $|+\rangle = \{ \tikzcircle[black, fill=white]{5pt}, \tikzcircle[black, fill=black]{5pt} \}$ and
the calculation is immediate following the approach shown in Section 
3 (or, more laboriously, by translating it to Dirac notation and 
then working
out the algebra):
\begin{equation*} \label{eq12}
\begin{split}
 & \{\{\{\tikzcircle[black, fill=white]{5pt}, \tikzcircle[black, fill=black]{5pt}\}, \tikzcircle[black, fill=black]{5pt}\}~, \{\tikzcircle[black, fill=white]{5pt}, \{\tikzcircle[black, fill=white]{5pt},  \tikzcircle[black, fill=black]{5pt}\}\}\} = \\
= &  \{\{~~e^{i\cdot\frac{\pi}{4}}~, e^{i\cdot\frac{\pi}{2}} \}~~, \{e^{i\cdot0}, ~~e^{i\cdot\frac{\pi}{4}}~~\}\} = \\ 
= &  \{~~~~~~~e^{i\cdot\frac{3\pi}{8}}~~~~~~, ~~~~~~~e^{i\cdot\frac{\pi}{8}}~~~~~~\} = \\ 
= &  ~~~~~~~~~~~~~~~~~~~e^{i\cdot\frac{\pi}{4}} ~~~~~~~ = \{\tikzcircle[black, fill=white]{5pt}, \tikzcircle[black, fill=black]{5pt}\}
\end{split}
\end{equation*}
The formalism introduced in \cite{b2} cannot handle these transformations. However, the group of gates introduced in the book,  
and the associated formalism presented there, are 
indeed universal. 
This, then, is the story of how the misty state formalism  came to be: “A
few years ago I was in the middle of pondering [Shi's] result when I realized I was running late to give a talk at a math
camp for 12-14 year olds [...] run in part by my friend PETE Shadbolt. I raced for the tube, and while on it thought
about what could I explain to these kids that wasn’t the usual jargon-filled quantum fluff. And so here we are.” \cite{b5}

\section{Entanglement Swapping}
Consider the following circuit:
\begin{figure}[h!]
    \centering
    \includegraphics[width=0.83\textwidth]{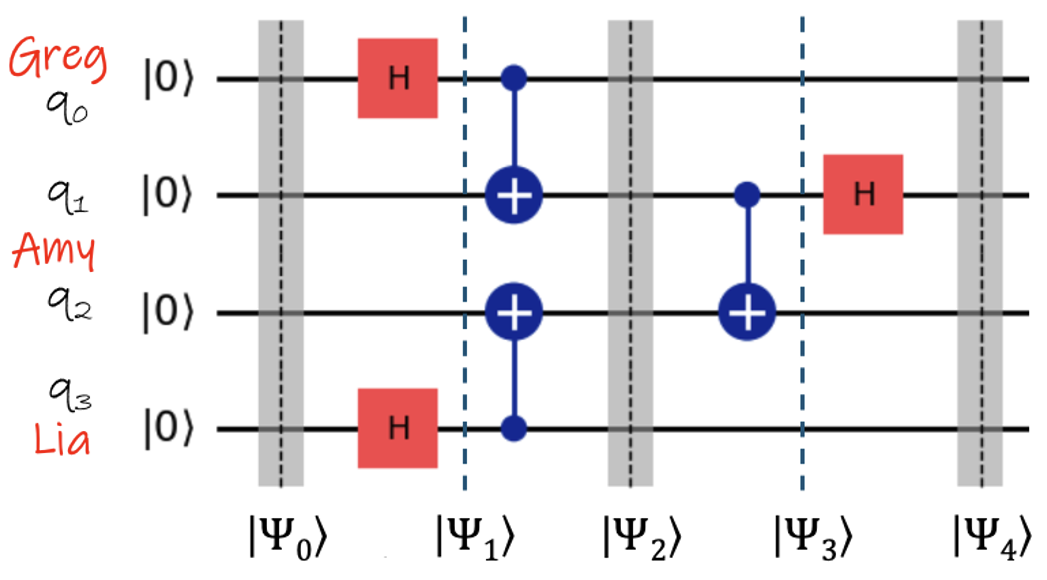}
    \caption{The circuit for entanglement swapping.}
    \label{fig:006}
\end{figure}

The basic setup for this circuit starts with Greg in Ohio, and Lia in the UK (Oxford). Amy is in California (Modesto) and 
she prepares two EPR pairs. Greg goes to Hawaii and stops in California, takes with him one half ($q_0$) of an EPR pair. Lia, too, 
is visiting from the UK, picks up a half of the other pair ($q_3$) and returns to Oxford (UK) with it. Amy measures her two qubits ($q_1$ 
and $q_2$) in the Bell basis and as a result Greg's and Lia's qubits ($q_0$ and $q_3$) become entangled. 

We're going to prove the above statement using misty states. To be sure, in three of the four possible cases Amy will need to look at her 
measurements and depending on what she sees she will have to call Greg and/or Lia and ask them to apply conditional correction gates. Here's how the proof goes:

$$|\Psi_0\rangle = \tikzcircle[black, fill=white]{5pt}\tikzcircle[black, fill=white]{5pt}\tikzcircle[black, fill=white]{5pt}\tikzcircle[black, fill=white]{5pt}$$

Through the action of the Hadamard gates on $q_0$ and $q_3$:

$$|\Psi_1\rangle = \{ \tikzcircle[black, fill=white]{5pt}, \tikzcircle[black, fill=black]{5pt} \}\tikzcircle[black, fill=white]{5pt}\tikzcircle[black, fill=white]{5pt} \{ \tikzcircle[black, fill=white]{5pt}, \tikzcircle[black, fill=black]{5pt}\} = \{ \tikzcircle[black, fill=white]{5pt}\tikzcircle[black, fill=white]{5pt}\tikzcircle[black, fill=white]{5pt}\tikzcircle[black, fill=white]{5pt}, \tikzcircle[black, fill=black]{5pt}\tikzcircle[black, fill=white]{5pt}\tikzcircle[black, fill=white]{5pt}\tikzcircle[black, fill=white]{5pt}, \tikzcircle[black, fill=white]{5pt}\tikzcircle[black, fill=white]{5pt}\tikzcircle[black, fill=white]{5pt}\tikzcircle[black, fill=black]{5pt}, \tikzcircle[black, fill=black]{5pt}\tikzcircle[black, fill=white]{5pt}\tikzcircle[black, fill=white]{5pt}\tikzcircle[black, fill=black]{5pt} \}$$

By action  of $q_0$ on $q_1$ and $q_3$ on $q_2$ (both as part of C-NOT gates) we have:

$$|\Psi_2\rangle =  \{ \tikzcircle[black, fill=white]{5pt}\tikzcircle[black, fill=white]{5pt}\tikzcircle[black, fill=white]{5pt}\tikzcircle[black, fill=white]{5pt}, \tikzcircle[black, fill=black]{5pt}\tikzcircle[black, fill=black]{5pt}\tikzcircle[black, fill=white]{5pt}\tikzcircle[black, fill=white]{5pt}, \tikzcircle[black, fill=white]{5pt}\tikzcircle[black, fill=white]{5pt}\tikzcircle[black, fill=black]{5pt}\tikzcircle[black, fill=black]{5pt}, \tikzcircle[black, fill=black]{5pt}\tikzcircle[black, fill=black]{5pt}\tikzcircle[black, fill=black]{5pt}\tikzcircle[black, fill=black]{5pt} \}$$

Similarly $q_1$ acts as a control qubit on target $q_2$ to produce: 

$$|\Psi_3\rangle =  \{ \tikzcircle[black, fill=white]{5pt}\tikzcircle[black, fill=white]{5pt}\tikzcircle[black, fill=white]{5pt}\tikzcircle[black, fill=white]{5pt}, \tikzcircle[black, fill=black]{5pt}\tikzcircle[black, fill=black]{5pt}\tikzcircle[black, fill=black]{5pt}\tikzcircle[black, fill=white]{5pt}, \tikzcircle[black, fill=white]{5pt}\tikzcircle[black, fill=white]{5pt}\tikzcircle[black, fill=black]{5pt}\tikzcircle[black, fill=black]{5pt}, \tikzcircle[black, fill=black]{5pt}\tikzcircle[black, fill=black]{5pt}\tikzcircle[black, fill=white]{5pt}\tikzcircle[black, fill=black]{5pt} \}$$

And now via the action of the Hadamard gate on $q_1$ we reach the final stage:

\begin{equation*} \label{eq203}
\begin{split}
|\Psi_4\rangle = \{ & \tikzcircle[black, fill=white]{5pt} \{ \tikzcircle[black, fill=white]{5pt}_1, \tikzcircle[black, fill=black]{5pt}_2 \} \tikzcircle[black, fill=white]{5pt}\tikzcircle[black, fill=white]{5pt}, \tikzcircle[black, fill=black]{5pt}\{ \tikzcircle[black, fill=white]{5pt}_3, \overline{\tikzcircle[black, fill=black]{5pt}}_4 \} \tikzcircle[black, fill=black]{5pt}\tikzcircle[black, fill=white]{5pt}, \tikzcircle[black, fill=white]{5pt}\{ \tikzcircle[black, fill=white]{5pt}_5, \tikzcircle[black, fill=black]{5pt}_6 \} \tikzcircle[black, fill=black]{5pt}\tikzcircle[black, fill=black]{5pt}, \tikzcircle[black, fill=black]{5pt}\{ \tikzcircle[black, fill=white]{5pt}_7, \overline{ \tikzcircle[black, fill=black]{5pt}}_8 \} \tikzcircle[black, fill=white]{5pt}\tikzcircle[black, fill=black]{5pt} \} \\
 = \{ & \tikzcircle[black, fill=white]{5pt} \tikzcircle[black, fill=white]{5pt}_1 \tikzcircle[black, fill=white]{5pt}\tikzcircle[black, fill=white]{5pt}, \tikzcircle[black, fill=black]{5pt} \tikzcircle[black, fill=white]{5pt}_7 \tikzcircle[black, fill=white]{5pt}\tikzcircle[black, fill=black]{5pt}, {\color{lightgray}\texttt{ pairing the first and seventh terms }} \\
   & \tikzcircle[black, fill=white]{5pt} \tikzcircle[black, fill=white]{5pt}_5 \tikzcircle[black, fill=black]{5pt}\tikzcircle[black, fill=black]{5pt}, \tikzcircle[black, fill=black]{5pt} \tikzcircle[black, fill=white]{5pt}_3 \tikzcircle[black, fill=black]{5pt}\tikzcircle[black, fill=white]{5pt}, {\color{lightgray}\texttt{ then the fifth with the third terms }} \\
      & \tikzcircle[black, fill=white]{5pt} \tikzcircle[black, fill=black]{5pt}_2 \tikzcircle[black, fill=white]{5pt}\tikzcircle[black, fill=white]{5pt}, \overline{\tikzcircle[black, fill=black]{5pt}} \tikzcircle[black, fill=black]{5pt}_8 \tikzcircle[black, fill=white]{5pt}\tikzcircle[black, fill=black]{5pt}, {\color{lightgray}\texttt{ pairs are grouped by middle two qubits }} \\
         & \tikzcircle[black, fill=white]{5pt} \tikzcircle[black, fill=black]{5pt}_6 \tikzcircle[black, fill=black]{5pt}\tikzcircle[black, fill=black]{5pt}, \overline{\tikzcircle[black, fill=black]{5pt}} \tikzcircle[black, fill=black]{5pt}_4 \tikzcircle[black, fill=black]{5pt}\tikzcircle[black, fill=white]{5pt} \} {\color{lightgray}\texttt{ with phase moved on the first qubit }}\\  
 = \{ & \_ \tikzcircle[black, fill=white]{5pt}\tikzcircle[black, fill=white]{5pt}\_ \{ \tikzcircle[black, fill=white]{5pt} \_ \_ \tikzcircle[black, fill=white]{5pt}, \tikzcircle[black, fill=black]{5pt} \_ \_ \tikzcircle[black, fill=black]{5pt} \},  {\color{lightgray}\texttt{ no correction needed }}  \\
 & \_ \tikzcircle[black, fill=white]{5pt}\tikzcircle[black, fill=black]{5pt}\_ \{ \tikzcircle[black, fill=white]{5pt} \_ \_ \tikzcircle[black, fill=black]{5pt}, \tikzcircle[black, fill=black]{5pt} \_ \_ \tikzcircle[black, fill=white]{5pt} \},  {\color{lightgray}\texttt{ Lia must apply X gate}}  \\
 & \_ \tikzcircle[black, fill=black]{5pt}\tikzcircle[black, fill=white]{5pt}\_ \{ \tikzcircle[black, fill=white]{5pt} \_ \_ \tikzcircle[black, fill=white]{5pt}, \overline{\tikzcircle[black, fill=black]{5pt}} \_ \_ \tikzcircle[black, fill=black]{5pt} \},  {\color{lightgray}\texttt{ Greg must apply Z gate }}  \\
 & \_ \tikzcircle[black, fill=black]{5pt}\tikzcircle[black, fill=black]{5pt}\_ \{ \tikzcircle[black, fill=white]{5pt} \_ \_ \tikzcircle[black, fill=black]{5pt}, \overline{\tikzcircle[black, fill=black]{5pt}} \_ \_ \tikzcircle[black, fill=white]{5pt} \} \}  {\color{lightgray}\texttt{ Greg applies Z, Lia X}}  
\end{split} 
\end{equation*}

And that concludes the analysis. 

\section{The GHZ Game}

The development of quantum technologies has seen a tremendous
upsurge in recent years and the theory of Bell nonlocality has been
key in making these technologies possible. Entanglement is a key
resource in quantum cryptography and quantum computing. Entanglement is capable of superluminal correlations. 

The GHZ game
is a famous thought experiment in quantum mechanics and computer science that demonstrates quantum pseudo-telepathy. It is a
collaborative, three-player, one-round game where players can win
100\% of the time using quantum entanglement, whereas the best
classical strategy only wins 75\% of the time. Besides
having fundamental implications, nonlocality is so specific that
it can be used to develop and certify reliable quantum devices.

Following \cite{b32} we will summarize the GHZ game as follows: there are three players, Alice, Bob and Carol playing against a referee. The referee
draws a triplet $(x, y, z)$ randomly from the four options shown in the table below and listed here for your convenience: $(0, 0, 0), (1, 1, 0), (1, 0, 1)$ and 
$(0, 1, 1)$. Alice, Bob and Carol each respond with an answer $a$, $b$ and $c$ also in the form of $0$  or $1$. The players can formulate a strategy prior to the
start of the game. However, no communication is allowed during the game itself.

\begin{center}
\begin{tabular}{ c|c|c|c } 
\multicolumn{4}{c}{Winning condition of GHZ game} \\
 \hline
 $x$ & $y$ & $z$ & $a + b + c$ \\ 
 \hline 
 \hline 
 $0$ & $0$ & $0$ & $0$ mod $2$ \\ 
 \hline 
 $1$ & $1$ & $0$ & $1$ mod $2$ \\ 
 $1$ & $0$ & $1$ & $1$ mod $2$ \\ 
 $0$ & $1$ & $1$ & $1$ mod $2$ \\ 
 \hline
\end{tabular}
\end{center}

\noindent Players win if the sum of their answers $a + b + c$ is even when $x = y = z = 0$ and odd in the other three cases. It can be shown \cite{b33} that there
is no classical strategy that satisfies all four winning conditions simultaneously and that, in fact, the best they can do is to  win 75\% of the time. 

However, if they are allowed 
to share a tripartite entangled state (known as the GHZ state) $|\psi\rangle = \frac{1}{\sqrt{2}}(|000\rangle + |111\rangle)$ they can win all the time (i.e., with probability 1).
The strategy they need to adopt is as follows: any player that receives a $0$ must make a measurement in the $X$ basis; any player 
receiving a $1$ must make a measurement in the $Y$ basis. If the measurement comes out a $|+\rangle$ or a $|\!+\!i\rangle$ 
the player
responds with a $0$. Otherwise the player outputs a $1$. 

The $X$ basis is made of the eigenstates
of the Pauli $X$ operator: $$|+\rangle = \frac{1}{\sqrt{2}}\big(|0\rangle + |1\rangle\big)~\textrm{and}~|-\rangle = \frac{1}{\sqrt{2}}\big(|0\rangle - |1\rangle\big)$$
The $Y$ basis is made of the eigenstates
of the Pauli $Y$ operator: $$|\!+\!i\rangle = \frac{1}{\sqrt{2}}\big(|0\rangle + i|1\rangle\big)~\textrm{and}~|\!-\!i\rangle = \frac{1}{\sqrt{2}}\big(|0\rangle - i|1\rangle\big)$$
The next section shows the calculation. In  the spirit of \cite{b2} we assume the reader sees $i$ here 
for the first time. In the spirit of \cite{fie2023} we provide, here and there,
milestones expressed in the traditional algebraic formalism and equivalents of the calculations 
done using misty states, for reinforcement and as a reality check.

As a reminder, eigenvectors are just fixed points.

\subsection{The Calculations}

The $Z$-basis is made of $\tikzcircle[black, fill=white]{5pt} = |0\rangle = \begin{pmatrix} 1 \\
0
\end{pmatrix}$ and $\tikzcircle[black, fill=black]{5pt} = |1\rangle = \begin{pmatrix} 0 \\
1
\end{pmatrix}$. 

The $X$-basis is made of the eigenstates of the Pauli $X$ operator defined through $X(\tikzcircle[black, fill=white]{5pt}) = \tikzcircle[black, fill=black]{5pt}$ and 
$X(\tikzcircle[black, fill=black]{5pt}) = \tikzcircle[black, fill=white]{5pt}$. As a reminder, these two axioms effectively extract the two columns of $X$  
represented through its unitary: $
\begin{pmatrix}
    0 & 1 \\
    1 & 0
\end{pmatrix}$. 

The eigenstates of $X$ are $$|+\rangle = H(\tikzcircle[black, fill=white]{5pt}) = \{ \tikzcircle[black, fill=white]{5pt}, \tikzcircle[black, fill=black]{5pt} \} \stackrel{\text{def}}{=} \tikzcircle[black, fill=yellow]{5pt}~~\textrm{with}~~X(\tikzcircle[black, fill=yellow]{5pt}) = \{ \tikzcircle[black, fill=black]{5pt}, \tikzcircle[black, fill=white]{5pt} \} = \tikzcircle[black, fill=yellow]{5pt}$$ and $$|-\rangle = H(\tikzcircle[black, fill=black]{5pt}) = \{ \tikzcircle[black, fill=white]{5pt}, \overline{\tikzcircle[black, fill=black]{5pt}} \} \stackrel{\text{def}}{=} \tikzcircle[black, fill=red]{5pt}~~\textrm{where}~~X(\tikzcircle[black, fill=red]{5pt}) = \{ \tikzcircle[black, fill=black]{5pt}, \overline{\tikzcircle[black, fill=white]{5pt}} \} = - \tikzcircle[black, fill=red]{5pt} = \overline{\tikzcircle[black, fill=red]{5pt}}$$
From these we deduce $\tikzcircle[black, fill=white]{5pt} = \{ \tikzcircle[black, fill=yellow]{5pt}, \tikzcircle[black, fill=red]{5pt} \}$ and 
$\tikzcircle[black, fill=black]{5pt} = \{ \tikzcircle[black, fill=yellow]{5pt}, \overline{\tikzcircle[black, fill=red]{5pt}} \}$.

The $Y$-basis is made of the eigenstates of the Pauli $Y$ operator which is defined via 
$$Y(\tikzcircle[black, fill=white]{5pt}) = i \tikzcircle[black, fill=black]{5pt}$$ and 
$$Y(\tikzcircle[black, fill=black]{5pt}) = - i \tikzcircle[black, fill=white]{5pt} = i \overline{ \tikzcircle[black, fill=white]{5pt}}$$ These axioms effectively extract the two columns of $Y=
\begin{pmatrix}
    0 & -i \\
    i & 0
\end{pmatrix}$ and $i = \sqrt{-1}$. 

The eigenstates of $Y$ are $$|+i\rangle =  \{ \tikzcircle[black, fill=white]{5pt}, i \tikzcircle[black, fill=black]{5pt} \} \stackrel{\text{def}}{=} \tikzcircle[black, fill=lime]{5pt}~~\textrm{with}~~Y(\tikzcircle[black, fill=lime]{5pt}) = \{ Y(\tikzcircle[black, fill=white]{5pt}), iY(\tikzcircle[black, fill=black]{5pt}) \} = \{ i \tikzcircle[black, fill=black]{5pt}, \tikzcircle[black, fill=white]{5pt} \} = \tikzcircle[black, fill=lime]{5pt}$$ and $$|-i\rangle =  \{ \tikzcircle[black, fill=white]{5pt}, i \overline{\tikzcircle[black, fill=black]{5pt}} \} \stackrel{\text{def}}{=} \tikzcircle[black, fill=pink]{5pt}~~\textrm{where}~~Y(\tikzcircle[black, fill=pink]{5pt}) = \{ i \tikzcircle[black, fill=black]{5pt}, i^2 \tikzcircle[black, fill=white]{5pt} \} = - \tikzcircle[black, fill=pink]{5pt} = \overline{\tikzcircle[black, fill=pink]{5pt}}$$
From these we deduce $\tikzcircle[black, fill=white]{5pt} = \{ \tikzcircle[black, fill=lime]{5pt}, \tikzcircle[black, fill=pink]{5pt} \}$ and 
$\tikzcircle[black, fill=black]{5pt} = \{ i \overline{ \tikzcircle[black, fill=lime]{5pt}}, i \tikzcircle[black, fill=pink]{5pt} \}$.

\subsection{Playing the Game}

There are two cases. In the first case the players all receive inputs $0$. In that case they all have to measure in the $X$ basis. The tripartite entangled state becomes:
$$\{\tikzcircle[black, fill=white]{5pt}\tikzcircle[black, fill=white]{5pt}\tikzcircle[black, fill=white]{5pt}, 
    \tikzcircle[black, fill=black]{5pt}\tikzcircle[black, fill=black]{5pt}\tikzcircle[black, fill=black]{5pt}\} = 
    \{ \{ \tikzcircle[black, fill=yellow]{5pt}, \tikzcircle[black, fill=red]{5pt} \}\{ \tikzcircle[black, fill=yellow]{5pt}, \tikzcircle[black, fill=red]{5pt} \}\{ \tikzcircle[black, fill=yellow]{5pt}, \tikzcircle[black, fill=red]{5pt} \}, \{ \tikzcircle[black, fill=yellow]{5pt}, \overline{\tikzcircle[black, fill=red]{5pt}} \} \{ \tikzcircle[black, fill=yellow]{5pt}, \overline{\tikzcircle[black, fill=red]{5pt}} \} \{ \tikzcircle[black, fill=yellow]{5pt}, \overline{\tikzcircle[black, fill=red]{5pt}} \}  \}$$
This then becomes:
\begin{align*}
\{\tikzcircle[black, fill=white]{5pt}\tikzcircle[black, fill=white]{5pt}\tikzcircle[black, fill=white]{5pt}, 
    \tikzcircle[black, fill=black]{5pt}\tikzcircle[black, fill=black]{5pt}\tikzcircle[black, fill=black]{5pt}\} =  \{ & \tikzcircle[black, fill=yellow]{5pt}\tikzcircle[black, fill=yellow]{5pt}\tikzcircle[black, fill=yellow]{5pt},
\tikzcircle[black, fill=yellow]{5pt}\tikzcircle[black, fill=yellow]{5pt}\tikzcircle[black, fill=red]{5pt},
\tikzcircle[black, fill=yellow]{5pt}\tikzcircle[black, fill=red]{5pt}\tikzcircle[black, fill=yellow]{5pt},
\tikzcircle[black, fill=yellow]{5pt}\tikzcircle[black, fill=red]{5pt}\tikzcircle[black, fill=red]{5pt},
\tikzcircle[black, fill=red]{5pt}\tikzcircle[black, fill=yellow]{5pt}\tikzcircle[black, fill=yellow]{5pt},
\tikzcircle[black, fill=red]{5pt}\tikzcircle[black, fill=yellow]{5pt}\tikzcircle[black, fill=red]{5pt},
\tikzcircle[black, fill=red]{5pt}\tikzcircle[black, fill=red]{5pt}\tikzcircle[black, fill=yellow]{5pt},
\tikzcircle[black, fill=red]{5pt}\tikzcircle[black, fill=red]{5pt}\tikzcircle[black, fill=red]{5pt}, \\
 & \tikzcircle[black, fill=yellow]{5pt}\tikzcircle[black, fill=yellow]{5pt}\tikzcircle[black, fill=yellow]{5pt},
\tikzcircle[black, fill=yellow]{5pt}\tikzcircle[black, fill=yellow]{5pt}\overline{\tikzcircle[black, fill=red]{5pt}},
\tikzcircle[black, fill=yellow]{5pt}\overline{\tikzcircle[black, fill=red]{5pt}}\tikzcircle[black, fill=yellow]{5pt},
\tikzcircle[black, fill=yellow]{5pt}\overline{\tikzcircle[black, fill=red]{5pt}}\overline{\tikzcircle[black, fill=red]{5pt}},
\overline{\tikzcircle[black, fill=red]{5pt}}\tikzcircle[black, fill=yellow]{5pt}\tikzcircle[black, fill=yellow]{5pt},
\overline{\tikzcircle[black, fill=red]{5pt}}\tikzcircle[black, fill=yellow]{5pt}\overline{\tikzcircle[black, fill=red]{5pt}},
\overline{\tikzcircle[black, fill=red]{5pt}}\overline{\tikzcircle[black, fill=red]{5pt}}\tikzcircle[black, fill=yellow]{5pt},
\overline{\tikzcircle[black, fill=red]{5pt}}\overline{\tikzcircle[black, fill=red]{5pt}}\overline{\tikzcircle[black, fill=red]{5pt}} \}  = \\ 
 = \{ & \tikzcircle[black, fill=yellow]{5pt}\tikzcircle[black, fill=yellow]{5pt}\tikzcircle[black, fill=yellow]{5pt},
\tikzcircle[white, fill=white]{5pt}\tikzcircle[white, fill=white]{5pt}\tikzcircle[white, fill=white]{5pt}~\,
\tikzcircle[white, fill=white]{5pt}\tikzcircle[white, fill=white]{5pt}\tikzcircle[white, fill=white]{5pt}~\,
\tikzcircle[black, fill=yellow]{5pt}\tikzcircle[black, fill=red]{5pt}\tikzcircle[black, fill=red]{5pt},
\tikzcircle[white, fill=white]{5pt}\tikzcircle[white, fill=white]{5pt}\tikzcircle[white, fill=white]{5pt}~
\tikzcircle[black, fill=red]{5pt}\tikzcircle[black, fill=yellow]{5pt}\tikzcircle[black, fill=red]{5pt},
\tikzcircle[black, fill=red]{5pt}\tikzcircle[black, fill=red]{5pt}\tikzcircle[black, fill=yellow]{5pt}~\,
\tikzcircle[white, fill=white]{5pt}\tikzcircle[white, fill=white]{5pt}\tikzcircle[white, fill=white]{5pt} \} 
\end{align*}
If you go back to the section in which we formulated the quantum strategy you will find out that we said the following:
if the players measure and obtain a red\footnote{Now that we have the colors.} (or pink) outcome they need to output a $1$ otherwise 
(for yellow or lime) they output a $0$. And we see
above that triplets with an odd number of red outcomes cancel each other due to destructive interference. 

What is left is
a set of triplets with an even number of red outcomes. The players then will output a sum $a + b + c$ that is even, so in  
this case they always win (regardless of their actual results obtained when they measure, all possible outcomes are listed above).

In the other three cases two of the players receive a $1$ and the third one receives a $0$ from the referee. The players that receive a $1$ need to measure in the
$Y$ basis, the third one needs to measure in the $X$ basis. We will calculate what happens in only one of the three cases\footnote{And claim the same result in the other
two, by symmetry.}, e.g. $(1, 1, 0)$. So Alice and Bob need to measure
in the $Y$ basis and Carol in the X basis. At the end we need to look into the resulting set of outcomes and replace red and pink with $1$ and replace yellow and lime with $0$
and calculate $a + b + c$ for each outcome to determine if the players win or not. The tripartite entangled state becomes:
$$\{\tikzcircle[black, fill=white]{5pt}\tikzcircle[black, fill=white]{5pt}\tikzcircle[black, fill=white]{5pt}, 
    \tikzcircle[black, fill=black]{5pt}\tikzcircle[black, fill=black]{5pt}\tikzcircle[black, fill=black]{5pt}\} = 
    \{ \{ \tikzcircle[black, fill=lime]{5pt}, \tikzcircle[black, fill=pink]{5pt} \}\{ \tikzcircle[black, fill=lime]{5pt}, \tikzcircle[black, fill=pink]{5pt} \}\{ \tikzcircle[black, fill=yellow]{5pt}, \tikzcircle[black, fill=red]{5pt} \}, \{ i\overline{\tikzcircle[black, fill=lime]{5pt}}, i \tikzcircle[black, fill=pink]{5pt} \}\{ i\overline{\tikzcircle[black, fill=lime]{5pt}}, i \tikzcircle[black, fill=pink]{5pt} \} \{ \tikzcircle[black, fill=yellow]{5pt}, \overline{\tikzcircle[black, fill=red]{5pt}} \}  \}$$
Performing the same calculation steps as before we have:
\begin{align*}
\{\tikzcircle[black, fill=white]{5pt}\tikzcircle[black, fill=white]{5pt}\tikzcircle[black, fill=white]{5pt}, 
    \tikzcircle[black, fill=black]{5pt}\tikzcircle[black, fill=black]{5pt}\tikzcircle[black, fill=black]{5pt}\} =  \{ & \tikzcircle[black, fill=lime]{5pt}\tikzcircle[black, fill=lime]{5pt}\tikzcircle[black, fill=yellow]{5pt},
\tikzcircle[black, fill=lime]{5pt}\tikzcircle[black, fill=lime]{5pt}\tikzcircle[black, fill=red]{5pt},
\tikzcircle[black, fill=lime]{5pt}\tikzcircle[black, fill=pink]{5pt}\tikzcircle[black, fill=yellow]{5pt},
\tikzcircle[black, fill=lime]{5pt}\tikzcircle[black, fill=pink]{5pt}\tikzcircle[black, fill=red]{5pt},
\tikzcircle[black, fill=pink]{5pt}\tikzcircle[black, fill=lime]{5pt}\tikzcircle[black, fill=yellow]{5pt},
\tikzcircle[black, fill=pink]{5pt}\tikzcircle[black, fill=lime]{5pt}\tikzcircle[black, fill=red]{5pt},
\tikzcircle[black, fill=pink]{5pt}\tikzcircle[black, fill=pink]{5pt}\tikzcircle[black, fill=yellow]{5pt},
\tikzcircle[black, fill=pink]{5pt}\tikzcircle[black, fill=pink]{5pt}\tikzcircle[black, fill=red]{5pt}, \\
 &\overline{\tikzcircle[black, fill=lime]{5pt}}\overline{\tikzcircle[black, fill=lime]{5pt}}\overline{\tikzcircle[black, fill=yellow]{5pt}},
\overline{\tikzcircle[black, fill=lime]{5pt}}\overline{\tikzcircle[black, fill=lime]{5pt}}\overline{\overline{\tikzcircle[black, fill=red]{5pt}}},
\overline{\tikzcircle[black, fill=lime]{5pt}}\tikzcircle[black, fill=pink]{5pt}\overline{\tikzcircle[black, fill=yellow]{5pt}},
\overline{\tikzcircle[black, fill=lime]{5pt}}\tikzcircle[black, fill=pink]{5pt}\overline{\overline{\tikzcircle[black, fill=red]{5pt}}},
\tikzcircle[black, fill=pink]{5pt}\overline{\tikzcircle[black, fill=lime]{5pt}}\overline{\tikzcircle[black, fill=yellow]{5pt}},
\tikzcircle[black, fill=pink]{5pt}\overline{\tikzcircle[black, fill=lime]{5pt}}\overline{\overline{\tikzcircle[black, fill=red]{5pt}}},
\tikzcircle[black, fill=pink]{5pt}\tikzcircle[black, fill=pink]{5pt}\overline{\tikzcircle[black, fill=yellow]{5pt}},
\tikzcircle[black, fill=pink]{5pt}\tikzcircle[black, fill=pink]{5pt}\overline{\overline{\tikzcircle[black, fill=red]{5pt}}} \}  \\ 
\end{align*}
In calculating the second line of outcome triplets above we have used the fact that $i$ only shows with the second power, so it transforms in  a unary minus. We
placed that as an extra phase on Carol's outcome (third in each triplet). 

You see then that half of the resulting triplets cancel each other and we are left with:
$$\{\tikzcircle[black, fill=white]{5pt}\tikzcircle[black, fill=white]{5pt}\tikzcircle[black, fill=white]{5pt}, 
    \tikzcircle[black, fill=black]{5pt}\tikzcircle[black, fill=black]{5pt}\tikzcircle[black, fill=black]{5pt}\} =  \{ 
\tikzcircle[black, fill=lime]{5pt}\tikzcircle[black, fill=lime]{5pt}\tikzcircle[black, fill=red]{5pt},
\tikzcircle[black, fill=lime]{5pt}\tikzcircle[black, fill=pink]{5pt}\tikzcircle[black, fill=yellow]{5pt},
\tikzcircle[black, fill=pink]{5pt}\tikzcircle[black, fill=lime]{5pt}\tikzcircle[black, fill=yellow]{5pt},
\tikzcircle[black, fill=pink]{5pt}\tikzcircle[black, fill=pink]{5pt}\tikzcircle[black, fill=red]{5pt} 
 \} $$

\section{Commencement}
Quantum physics is known to be challenging for two reasons: it describes counter-intuitive phenomena and employs rather advanced mathematics. However, as we have shown, 
it is possible for a beginner not versed in mathematics to tackle mind-boggling, bona fide quantum experiments like entanglement swapping and the violation of Bell’s 
inequalities, and practice notions like superposition, interference, entanglement and measurement using some of the formalisms developed specifically for that purpose 
within the last 10-12 years.

\subsubsection{\discintname}
The author has no competing interests to declare that are
relevant to the content of this article. 

%
%
%

\end{document}